\documentclass[11pt]{revtex4-1}
\usepackage{natbib}
\usepackage{amssymb,amsmath,xcolor,graphicx,xspace,colortbl,textcomp, rotating}

\usepackage{color}

\begin{document}

\title{Topological invariants and the definition of energy in quadratic gravity theory}

\author{Gaston Giribet}
\affiliation{Physics Department, University of Buenos Aires \& IFIBA-CONICET,
Ciudad Universitaria, pabell\'on 1, 1428, Buenos Aires, Argentina.}
\email{gaston@df.uba.ar}
\author{Olivera Miskovic}
\affiliation{Instituto de F\'{\i}sica, Pontificia Universidad Cat\'olica de Valpara\'{\i}so, Casilla 4059, Valpara\'{\i}so, Chile.}
\email{olivera.miskovic@pucv.cl}
\author{Rodrigo Olea}
\affiliation{Departamento de Ciencias F\'{\i}sicas, Universidad Andres Bello, Sazi\'e 2212, Piso 7, Santiago, Chile.}
\email{rodrigo.olea@unab.cl}
\author{David Rivera-Betancour}
\affiliation{CPHT- Centre de Physique Th\'eorique, CNRS, \'{E}cole Polytechnique, Institut Polytechnique de Paris, Route de Saclay, 91128 Palaiseau, France.}
\email{david.rivera-betancour@zimbra.polytechnique.fr}

\begin{abstract}
\vspace{1cm}

We provide a simple method to compute the energy in higher curvature gravity in asymptotically AdS spacetimes in even dimensions. It follows from the combined use of topological terms added to the gravity action, and the Wald charges derived from the augmented action functional. No additional boundary terms are needed. As a consistency check, we show that the formula for the conserved quantities derived in this way yields the correct result for the mass of asymptotically AdS black holes.
\end{abstract}

\maketitle

\section{Introduction}

%%%%%%%%%%%%%%%%%%%%%%%%%%%%%%%%%%%%%%%%%%%

Higher-curvature corrections to Einstein-Hilbert action are ubiquitous in effective field theory when gravity is involved. In fact, a sensible theory of quantum gravity is generically expected to yield such corrections, and string theory, as antonomastic example of this, does predict them at next-to-leading order \cite{Zwiebach:1985uq, Gross:1986iv, {Metsaev:1987zx}}. Higher-curvature corrections, on the other hand, are interesting by their own right: they have been studied since long time ago and in many different contexts, including mathematical aspects of general relativity (GR) \cite{Lanczos:1938sf, Lovelock:1971yv, Lovelock:1972vz}, cosmology \cite{Starobinsky:1980te}, black holes \cite{Boulware:1985wk}, massive gravity \cite{Stelle2}, supergravity \cite{Ferrara:1977mv}, and quantum gravity \cite{Stelle}.

Regarding quantum gravity, it is well-known that the introduction of higher-order terms in the gravity action suffices to render the theory renormalizable \cite{Stelle} but at the price of introducing ghosts \cite{Stelle2}. This is generically the case, with few notable exceptions \cite{Zwiebach:1985uq}. Therefore, the absence of ghosts and of other potential pathologies such as causality issues, can be used as a criterion to select the sensible higher-order theory or ultraviolet completion to work with \cite{Zwiebach:1985uq, Gruzinov:2006ie, Camanho:2014apa}.

%%%%%%%%%%%%%%%%%%%%%%%%%%%%%%%%%%%%%%%%%%%%

In the recent years, the interest in {\it healthy} higher-order corrections to Einstein theory has been renewed. In particular, there have been very interesting works studying higher-curvature models in anti-de Sitter (AdS) space. Such is the case of the so-called Critical Gravity (CG) theories \cite{Lu:Pope, Deser:2011xc}, which provide ghost-free models of gravity in asymptotically AdS spacetimes in $D\geq 4$ dimensions. In $D=3$ dimensions, higher-curvature terms were also considered as a toy model of a consistent gravity theory \cite{Li:2008dq}. Other recent works explore, for instance, the connection between higher-curvature conformally invariant theories and Einstein gravity in (A)dS in $D=4$ \cite{Maldacena:2011mk}. Other models in $D\geq 4$ dimensions studied recently, such as the so-called (generalized) quasi-topological gravity \cite{Oliva:2010zd, Oliva:2010eb, Myers:2010ru,Hennigar:2017ego,Bueno:2017sui,Bueno:2019ycr} and the Einstenian gravity \cite{Bueno:2016xff, Arciniega:2018fxj} are also very interesting and provide a new perspective on this old topic.

%%%%%%%%%%%%%%%%%%%%%%%%%%%%%%%%%%%%%%%%%%%%%%

The computation of conserved charges in higher-curvature theories in both asymptotically flat and asymptotically AdS spaces is an important problem that has been addressed by many authors in the last twenty years; notably by Deser and Tekin, who found in Refs.\cite{Deser:BTekin, Deser:Tekin} the higher-derivative generalization of the Abbot-Deser method for GR \cite{Abbott:Deser}. Other papers discussing conserved charges in related contexts are \cite{Okuyama, Pang, Azeyanagi:2009wf}, and of course other using the covariant formalism of Ref.\cite{Barnich:2001jy}; see references thereof.

Recently, we proposed in \cite{GORD} a novel definition of gravitational energy for an arbitrary theory of gravity including quadratic-curvature corrections to Einstein equations. We focused on the theory in $D=4$ dimensions and in presence of negative cosmological constant. Unlike some other methods considered in the literature, the method proposed in \cite{GORD} is intrinsically non-linear and permits to deal with the boundary terms \cite{Iyer:Wald, Wald:Zoupas} in a systematic way. It relies on the idea of adding to the gravity action topological invariant terms, which suffice to regularize the Noether charges and render the variational problem well-posed. --More precisely, the method amounts to adding to the action the bulk part of a topological invariant.-- This is an idea that has been previously considered in \cite{Aros, Aros2} in the case of second-order theories, such as Einstein-AdS gravity. Here, we show how this method can be generalized to generic quadratic-curvature theories in any even dimension.

The paper is organized as follows: In section II, we will introduce the action of generic quadratic-curvature gravity in $D=2n$ dimensions, including topological invariants that will eventually serve to regularize the action. In section III, we provide the definition of gravitational energy and we use it to compute the mass of black holes in asymptotically AdS$_D$ space. Section IV contains our conclusions.

%The existence of Topological Invariants is of vital importance to produce
%important results in Mathematics as the Five-color Theorem \cite{Richeson}.
%In Physics, ground-breaking discoveries are the consequence of the inclusion
%of topological terms in the Lagrangian, even though they do not modify the
%field equations. In point of fact, if one believes that the description of a
%physical system is not only given by the equations of motion in the bulk,
%but also involves surface terms and adequate boundary conditions, there is
%no a priori reason to drop topological terms from the action. Clearly, the
%coupling of these topological terms cannot be determined from the bulk
%dynamics. However, an appropriate set of boundary conditions which, in turn,
%leads to a well-posed action principle, may be used for that purpose. In the
%case of 4D electromagnetism, the addition of a topological invariant
%(Pontryagin term) on top of the Maxwell action changes the junction
%conditions defined at the interface/boundary between different regions of
%the space. This explains the striking properties of materials as Topological
%%Insulators \cite{TopInsulators,Hasan}. Apart from the local modifications of
%the boundary dynamics, the Pontryagin term is responsible for the appearance
%of magnetic charge as a topological (global) property in Condensed Matter
%systems. \newline

\section{Quadratic-Curvature Gravity in $D=2n$ dimensions}

The most general gravity action which adds up quadratic corrections in the
curvature to GR in even dimensions is given by the
expression
\begin{equation}
I=\int\limits_{M}d^{2n}x\sqrt{-g}\left( \frac{R-2\Lambda _{0}}{\kappa }%
+\alpha R_{\nu }^{\mu }R_{\mu }^{\nu }+\beta R^{2}+\gamma \,GB\right)
+\alpha _{2n}\int\limits_{M}d^{2n}x\,\mathcal{E}_{2n}\,.  \label{action}
\end{equation}%
\newline
Here, $GB\equiv\sqrt{-g}\left(R^{\alpha \beta \mu \nu}R_{\alpha \beta \mu \nu} -4R^{\mu \nu}R_{\mu \nu}+R^2\right)$ is the Gauss-Bonnet term, which in $D>4$ is a dynamical term. The
last term in the integral (\ref{action}) is, by contrast, the $2n$-dimensional topological
Euler density, which reads
\begin{equation}
\mathcal{E}_{2n}=\frac{\sqrt{-g}}{2^{n}}\,\delta _{\lbrack \nu _{1}\cdots\nu
_{2n}]}^{[\mu _{1}\cdots\mu _{2n}]}R_{\mu _{1}\mu _{2}}^{\nu _{1}\nu
_{2}}\cdots R_{\mu _{2n-1}\mu _{2n}}^{\nu _{2n-1}\nu _{2n}}\,.  \label{Euler}
\end{equation}
This contribution accommodates the maximal number of curvatures in $2n$ dimension, such that it does not contribute to the field equations.
In $D=4$, this invariant is the same as the Gauss-Bonnet term.

We shall use the conventions in Ref.\cite{Deser:Tekin},
where $\kappa =2\text{Vol}(S^{2n-2})G_{2n}$ in terms of the $D$-dimensional Newton's
constant $G_{D}$. In addition, in order to compare with existing literature,
we are considering the inclusion of a bare cosmological constant $\Lambda _{0}=-(2n-1)(2n-2)/2\ell^{2}$ in terms of the original AdS radius $\ell$. In our notation, $\delta _{\lbrack \nu _{1}\cdots \nu _{p}]}^{[\mu _{1}\cdots \mu _{p}]}=\det \left[ \delta _{\nu _{1}}^{\mu _{1}}\cdots \delta
_{\nu _{p}}^{\mu _{p}}\right] $ is a totally antisymmetric product of $p$ Kronecker deltas.

Varying the action with respect to the metric, one obtains the equations of motion (EOM), which correspond to
Einstein tensor plus fourth-order contributions coming from
curvature-squared terms; namely
\begin{eqnarray}
0 &=&\frac{1}{\kappa }\,G_{\mu \nu }+2\beta R\left( R_{\mu \nu }-\frac{1}{4}\,g_{\mu \nu }R\right) +\left( 2\beta +\alpha \right) \left( g_{\mu \nu
}\Box -\nabla _{\mu }\nabla _{\nu }\right) R - \notag \\
&&-\gamma H_{\mu \nu }+\alpha \Box G_{\mu \nu }+2\alpha \left( R_{\mu \sigma \nu \rho }-\frac{1}{4}\,g_{\mu \nu }R_{\sigma \rho }\right) R^{\sigma \rho }\,,  \label{EOM}
\end{eqnarray}
where the part linear in the curvature,
\begin{equation}
G_{\nu }^{\mu }=R_{\nu }^{\mu }-\frac{1}{2}\,R\delta _{\nu }^{\mu }+\Lambda
_{0}\delta _{\nu }^{\mu }\,,
\end{equation}%
is the usual field equation of GR, and the quadratic part
\begin{equation}
H_{\nu }^{\mu }=\frac{1}{8}\,\delta _{\lbrack \nu \nu _{1}\nu _{2}\nu _{3}\nu
_{4}]}^{[\mu \mu _{1}\mu _{2}\mu _{3}\mu _{4}]}R_{\mu _{1}\mu _{2}}^{\nu
_{1}\nu _{2}}R_{\mu _{3}\mu _{4}}^{\nu _{3}\nu _{4}}\,,
\end{equation}%
corresponds to the Lanczos tensor \cite{Lanczos:1938sf, Lovelock:1971yv, Lovelock:1972vz}.

Vacuum states of the theory correspond to maximally-symmetric spaces, which
satisfy the constant curvature condition
\begin{equation}
R_{\alpha \beta }^{\mu \nu }=-\frac{1}{\ell _{\text{eff}}^{2}}\,\delta
_{\lbrack \alpha \beta ]}^{[\mu \nu ]}.  \label{MSS}
\end{equation}%
Here, $\ell _{\text{eff}}$ is the effective (A)dS radius. From the equation
of motion one can readily obtain an expression for the effective (A)dS
radius in terms of the couplings of the theory. Using
the standard relation between (A)dS length and the corresponding
cosmological constant $\Lambda _{\text{eff}}=-(2n-1)(2n-2)/2\ell _{\text{eff}%
}^{2}$, one arrives at the equation
\begin{equation}
-\frac{1}{2\kappa \Lambda _{\text{eff}}}+\frac{\Lambda_{0}}{2\kappa\Lambda^2_{\text{eff}}}=\frac{(2n-4)}{(2n-2)^{2}}\,(2n\beta +\alpha )+\gamma\,
\frac{(2n-4)(2n-3)}{(2n-2)(2n-1)}\,.  \label{cosmconst}
\end{equation}%
Note that $\Lambda _{\text{eff}}=\Lambda _{0}$ in four dimensions. This is no longer the case in $D>4$, as
higher curvature terms modify the effective cosmological constant.

The surface term of the theory arises from integrating by parts the
variation of the gravitational action in order to construct the EOMs.
Without loss of generality, one can always split such contribution in the pieces
that contain $\delta g$ and $\delta \Gamma $, respectively. They can be
easily derived employing Wald construction of Noether conserved quantities
in gravity theories \cite{Iyer:Wald}. Then, the surface term is cast in the
form
\begin{equation}
\Theta ^{\alpha }(\delta g,\delta \Gamma )=2E_{\mu \nu }^{\alpha \beta
}\,g^{\nu \lambda }\delta \Gamma _{\beta \lambda }^{\mu }+2\nabla ^{\mu
}E_{\mu \nu }^{\alpha \beta }\left( g^{-1}\delta g\right) _{\beta }^{\nu }\,,
\label{surface}
\end{equation}%
where the tensor $E_{\mu \nu }^{\alpha \beta }={\delta \mathcal{L}}/{\delta R_{\alpha
\beta }^{\mu \nu }}$ is the derivative of the
Lagrangian density with respect to the Riemann tensor which, for QCG theory, yields
\begin{equation}
E_{\mu \nu }^{\alpha \beta }=\frac{1}{2\kappa }\delta _{\lbrack \mu
\nu ]}^{[\alpha \beta ]}+\frac{1}{2}\alpha R_{[\mu }^{[\alpha }\delta _{\nu
]}^{\beta ]}+\beta R\delta _{\lbrack \mu \nu ]}^{[\alpha \beta ]}+\frac{1}{2}%
\gamma \delta _{\lbrack \mu \nu \nu _{3}\nu _{4}]}^{[\alpha \beta \mu
_{3}\mu _{4}]}R_{\mu _{3}\mu _{4}}^{\nu _{3}\nu _{4}}+\frac{n\alpha _{2n}}{%
2^{n}}\delta _{\lbrack \mu \nu \nu _{3}\cdots\nu _{2n}]}^{[\alpha \beta \mu
_{3}\cdots\mu _{2n}]}R_{\mu _{3}\mu _{4}}^{\nu _{3}\nu _{4}}\cdots R_{\mu
_{2n-1}\mu _{2n}}^{\nu _{2n}\nu _{2n}} \,.  \label{surfaceterm}
\end{equation}%

In what follows, we construct an energy definition for QCG in even
dimensions, as the natural generalization of the procedure shown in Ref.\cite{GORD}.

\section{Energy definition}

In the literature, there is a plethora of different approaches to deal with the general issue of defining energy for a given gravity theory \cite{Adami:2017phg}. Here, we shall apply Wald formalism \cite{Iyer:Wald, Wald:Zoupas}, as
the charges derived with such formalism for the case of quadratic theories take a relatively simple form: For a Lagrangian which is a function of the metric and the Riemann
curvature $\mathcal{L}(g_{\mu \nu },R_{\mu \nu \alpha \beta })$, the
conserved quantity associated to any Killing vector $\xi ^{\mu }$ is
expressed as a surface integral in the codimension-2 surface $\Sigma $
\begin{equation}
Q_{\text{W}}^{\alpha }[\xi ]=2\int\limits_{\Sigma }dS_{\beta }\left( E_{\mu \nu
}^{\alpha \beta }\nabla ^{\mu }\xi ^{\nu }+2\nabla ^{\mu }E_{\mu \nu
}^{\alpha \beta }\xi ^{\nu }\right) \,.  \label{NWcharge}
\end{equation}

The use of Wald's derivation of conserved quantities from the gravitational bulk Lagrangian
does not guarantee by itself that the value of the energy is correct. This can be seen
in GR, where Wald charge coincides with the Komar formula. So, even in the asymptotically
flat case, the black hole mass computed from (\ref{NWcharge}) does not coincide with
the one obtained by the Hamiltonian method. The situation worsens for
AdS gravity, as the behavior of solutions turns Komar integral divergent at the spatial infinity. These facts implies that the charges (\ref{NWcharge}) need to be corrected by adding suitable boundary terms
to the original Lagrangian. Prescribing appropriate boundary conditions makes possible the integration of the conserved quantities,
although getting closed expressions for the charges is not always guaranteed. In this regard, the covariant formalism \cite{Barnich:2001jy}, which provides a robust method to compute conserved charges in which the boundary terms are constructed systematically, is particularly useful. As we will see, we will obtain a result in agreement with that method.

The addition of topological terms to Einstein-Hilbert action in GR with negative cosmological constant provides a remarkably simple
method to circumvent the drawbacks in the procedure described above. In four dimensions, the consistent coupling of the Gauss-Bonnet and Pontryagin terms allows to express the gravitational charges as the
electric/magnetic part of the Weyl tensor \cite{Miskovic:Olea,Araneda:2016iiy}. The formulas, which give rise to the correct energy of different asymptotically AdS solutions, turn the discussions on background-substraction methods and the inclusion of extra boundary terms in the Lagrangian superfluous. In a previous work \cite{GORD}, we have shown that the addition of the
Gauss-Bonnet invariant to quadratic-curvature gravity (QCG) action in $D=4$
also acts as a regulator of the Noether charges in both Einstein and
non-Einstein sectors of the theory. Here, we extend this result to an arbitrary even-dimensional
QCG theory. A single topological term added to the action suffices to render the conserved
charges finite. The information about the background of the
corresponding sector of the theory is encoded in the asymptotic value of the curvature,
and gets reflected in the coupling constant of the Euler topological term. In other words, even in the higher-derivative theory it happens that the addition of topological invariants to the action is sufficient to regularize the conserved charges, making the job of the otherwise needed boundary terms.

In QCG, due to Bianchi identity, the second part of the  integrand in Eq.(\ref{NWcharge}) does
not feature any term containing $\gamma $. Altogether, the rest of $\nabla
^{\mu }E_{\mu \nu }^{\alpha \beta }$ vanishes identically for Einstein
spaces,
\begin{equation}
R_{\mu \nu }=-\frac{(2n-1)g_{\mu \nu }}{\ell _{\text{eff}}^{2}}\,.
\label{Einstein}
\end{equation}%
The gravity theory under analysis here admits a number of
analytic solutions with different asymptotic behavior, such as asymptotically AdS, Lifshitz, etc. These different asymptotic behaviors represent different sectors of the
theory. For the purpose of the present discussion, we will consider
solutions which are continuously connected to a global AdS spacetime, which
defines the background configuration. We then fix the coupling of the Euler density in terms of the parameters $%
\alpha $, $\beta $ and $\gamma $ by the following criterion: The total
surface term must vanish identically for the vacuum state (\ref{MSS})
corresponding to the class of solutions we are interested in. This basic
assumption implies a well-posed variational principle, at least for
global AdS spacetime. In doing so, the coupling constant of the Euler density reads
\begin{equation}
\alpha _{2n}=\frac{(-1)^{n}\ell _{\text{eff}}^{2n-2}}{n(2n-2)!\kappa }\left[
1+\frac{4\kappa \Lambda _{\text{eff}}}{2n-2}\left( 2n\beta +\alpha +\gamma
\frac{(2n-3)(2n-2)}{2n-1}\right) \right] \,.  \label{alppha}
\end{equation}%

The quartic relation for the effective AdS radius in (\ref{cosmconst})
has a single root provided the derivative of it with respect to $\ell_{\text{eff}}^{2}$ is different from zero. Indeed, this degeneracy condition poses an obstruction for the linearization of EOM when the two maximally symmetric vacua of the theory coincide. This is similar to what happens, e.g., in odd dimensions with the Chern-Simons gravity theories \cite{Arenas-Henriquez:2017xnr}. The construction
of conserved quantities is sensitive to this issue, as linearized charges cannot be obtained for the degenerate case.
In relation to this, it is worth emphasizing that the method of adding the Euler term (\ref{alppha}) is not affected by this consideration, as it does not rely on linear perturbations of the geometry around a given background. Once we have suitably identified the vacuum solution, we can ask whether the massive deviations from the maximally symmetric solution are such that the total surface integral keeps being finite. This can be checked in concrete examples.
\newline

The full charge is obtained as the surface integral
\begin{equation}
Q^{\alpha }[\xi ]=\int\limits_{\Sigma }dS_{\beta }\left( q_{(1)}^{\alpha
\beta }+q_{(2)}^{\alpha \beta }\right) \,,  \label{charge_top}
\end{equation}%
where $q_{(1)}^{\alpha \beta }$ and $q_{(2)}^{\alpha \beta }$ are the
prepotentials
\begin{equation*}
q_{(1)}^{\alpha \beta }=\nabla ^{\mu }\xi ^{\nu }\left( \frac{1}{\kappa }\,
\delta _{\lbrack \mu \nu ]}^{[\alpha \beta ]}+\alpha R_{[\mu }^{[\alpha
}\delta _{\nu ]}^{\beta ]}+2\beta R\delta _{\lbrack \mu \nu ]}^{[\alpha
\beta ]}+\gamma\, \delta _{\lbrack \mu \nu \nu _{3}\nu _{4}]}^{[\alpha \beta
\mu _{3}\mu _{4}]}R_{\mu _{3}\mu _{4}}^{\nu _{3}\nu _{4}}+\frac{n\alpha _{2n}%
}{2^{n-1}}\,\delta _{\lbrack \mu \nu \cdots \nu _{2_{n}}]}^{[\alpha \beta \cdots \mu
_{2_{n}}]}R_{\mu _{3}\mu _{4}}^{\nu _{3}\nu _{4}}\cdots R_{\mu _{2_{n-1}}\mu
_{2_{n}}}^{\nu _{2n-1}\nu _{2_{n}}}\right),
\end{equation*}%
and
\begin{equation}
q_{(2)}^{\alpha \beta }=2\nabla ^{\mu }\left( \alpha R_{[\mu }^{[\alpha
}\delta _{\nu ]}^{\beta ]}+2\beta R\delta _{\lbrack \mu \nu ]}^{[\alpha
\beta ]}\right) \xi ^{\nu }\,.
\end{equation}%

Now, we can consider static black hole with standard AdS asymptotics, whose metric is given by
\begin{equation}
ds^{2}=-f^{2}(r)\,dt^{2}+\frac{1}{f^{2}(r)}\,dr^{2}+r^{2}d\Omega _{2n-2}^{2}\,,
\label{blackhole}
\end{equation}%
where $d\Omega _{2n-2}^{2}$ is the metric on the unit $(2n-2)$-sphere. This discussion can be straightforwardly generalized to topological black holes with planar or hyperbolic horizons. The non-vanishing components of the Riemann curvature for this static configuration are
\begin{eqnarray}
R_{tr}^{tr} &=&-\frac{1}{2}\left( f^{2}\right) ^{\prime \prime }\,,\nonumber
\\
R_{tm}^{tn} &=&R_{rm}^{rn}=-\frac{1}{2r}\left( f^{2}\right) ^{\prime }\delta
_{m}^{n}\,, \nonumber\\
R_{kl}^{mn} &=& \frac{1-f^{2}}{r^{2}}\, \delta _{\lbrack
kl]}^{[mn]}\,,\label{Riemann}
\end{eqnarray}%
while the non-vanishing components of the Ricci tensor are
\begin{eqnarray}
R_{t}^{t} &=&R_{r}^{r}=-\frac{1}{2r}\left[ r\left( f^{2}\right) ^{\prime
\prime }+2(n-1)\left( f^{2}\right) ^{\prime }\right] \,,\nonumber \\
R_{m}^{n} &=&-\frac{1}{r^{2}}\left[ r\left( f^{2}\right) ^{\prime
}-(2n-3)\left( 1-f^{2}\right) \right] \delta _{m}^{n}\,.
\end{eqnarray}
Here, the prime stands for derivatives with respect to $r$. In the case of Boulware-Deser black holes \cite{Boulware:1985wk} of Einstein-Gauss-Bonnet gravity ($\alpha =\beta =0$), which is the working example in QCG, the metric
function $f^{2}(r)$ takes the asymptotic form
\begin{equation} \label{fallofff2}
f^{2}(r) \simeq \frac{r^{2}}{\ell _{%
\text{eff}}^{2}}+1-\left( \frac{r_{0}}{r}\right) ^{2n-3}+ \cdots\,,
\end{equation}%
with $\ell _{\text{eff}}^{2}$ being a root of the polynomial equation%
\begin{equation}
\gamma (2n-3)(2n-4)\,\frac{1}{\ell _{\text{eff}}^{4}}-\frac{1}{\kappa \ell _{%
\text{eff}}^{2}}+\frac{1}{\kappa \ell ^{2}}=0\,. \label{Leff}
\end{equation}
The ellipsis in (\ref{fallofff2}) stand for higher powers of $1/r$. In a more general case, in which the couplings $\alpha $ and $\beta $ are
non-zero and generic, a similar fall-off is expected, with the difference being the value of the AdS radius, which turns out to be given by equation (\ref{cosmconst}). For special relations between $\alpha $, $\beta $, and $\Lambda _0$, weakened version of the asymptotic condition (\ref{fallofff2}) might be possible. This has been extensively studied in the literature. However, for generic values of the parameters $\alpha $ and $\beta $, (\ref{fallofff2}) is the expected behavior, as the quadratic terms represent ultraviolet corrections. A more general case could eventually involve two different metric functions corresponding to $g_{tt}$ and $g^{-1}_{rr}$, each one with the asymptotic behavior that respects the AdS asymptotics as given by Eq.(\ref{fallofff2}), what we discuss at the end of this section.

For static massive objects, the energy formula takes the form
\begin{eqnarray}
E &=&Q^{t}[\partial _{t}]=\int\limits_{\Sigma }dS_{r}\nabla ^{r}\xi ^{t}
\left[ \alpha (R_{t}^{t}+R_{r}^{r})+\left( \frac{1}{\kappa }+2\beta R\right)
\delta _{\lbrack rt]}^{[rt]}\right. +  \notag \\
&&+\left. \gamma\, \delta _{\lbrack
rtm_{1}m_{2}]}^{[rtp_{1}p_{2}]}R_{p_{1}p_{2}}^{m_{1}m_{2}}+\frac{n\,\alpha
_{2n}}{2^{n-1}}\,\delta _{\lbrack
rtm_{1}\cdots m_{2n}]}^{[rtp_{1} \cdots p_{2n}]}R_{p_{1}p_{2}}^{m_{1}m_{2}}\cdots R_{p_{2n-3}p_{2n-2}}^{m_{2n-3}m_{2n-2}}%
 \rule{0pt}{15pt} \right].
\end{eqnarray}
Given the form of the line element for the black hole (\ref{blackhole}),
explicit evaluation of the above energy also requires the derivative of the time-like
Killing vector, $\nabla ^{r}\xi ^{t}=(f^2)'/2$, and the
codimension-2 surface element $dS_r$. If the local coordinates on $%
\Sigma $ are denoted by $y^{m}$ and the corresponding line element of $%
\Sigma $ is
\begin{equation}
r^{2}d\Omega _{2n-2}^{2}=r^{2}\sigma _{mn}(y)\,dy^{m}dy^{n},
\end{equation}
where $\sigma _{mn}$ is the metric of the unit sphere, then we find along the radial
normal  $dS_{r}=r^{2n-2}d^{2n-2}y\sqrt{\sigma }$. In turn, the
curvature expression in the square bracket can be recast in terms of the
metric function $f^{2}(r)$ and its derivatives. The result is
\begin{eqnarray}
E &=&\int\limits_{\Sigma }d^{2n-2}y\sqrt{\sigma } \,\frac{1}{2}\left(
f^{2}\right) ^{\prime }\left\{  \rule{0pt}{18pt} \frac{1}{\kappa }-\frac{\alpha }{r}\left(
r\left( f^{2}\right) ^{\prime \prime }+2(n-1)\left( f^{2}\right) ^{\prime
}\right) \right. -  \notag \\
&&-\frac{2\beta }{r^{2}}\left[ r^{2}\left( f^{2}\right) ^{\prime \prime
}+4(n-1)r\left( f^{2}\right) ^{\prime }-2(n-1)(2n-3)(1-f^{2})\right] +   \notag
\\
&&+\left. 4\gamma (n-1)(2n-3)\,\frac{1-f^{2}}{r^{2}} +n\,\alpha
_{2n}(2n-2)!\left( \frac{1-f^{2}}{r^{2}}\right) ^{n-1}\right\} \,.
\end{eqnarray}
 It is important to notice here that the Ricci tensor tends to a constant value, namely $R_{\nu }^{\mu }\simeq -(2n-1)\delta
_{\nu }^{\mu }/\ell _{\text{eff}}^{2}$.

Replacing the asymptotic value of the metric function and its derivatives and performing the integral on the angular variables $\int_{\Sigma }d^{2n-2}y\sqrt{\sigma }=$Vol$(S^{2n-2})$, the expression for the energy takes the form
\begin{eqnarray}
E &=&\text{Vol}(S^{2n-2})\lim_{r\rightarrow \infty }r^{2n-2}ff'\left[
\frac{1}{\kappa }-(\alpha +2\beta )\left( f^{2}\right) ^{\prime \prime
}-2(n-1)(\alpha +4\beta )\,\frac{\left( f^{2}\right) ^{\prime }}{r}+\right.  \nonumber \\  \label{EnergyLim}
&&\left.+4(n-1)(2n-3)\,\frac{1-f^{2}}{r^{2}}\,(\beta +\gamma )+n\,\alpha
_{2n}(2n-2)!\left( \frac{1-f^{2}}{r^{2}}\right) ^{n-1}\right] \, ,
\end{eqnarray}
where the last term of this expression can be expanded as
\begin{equation}
\left( \frac{1-f^{2}}{r^{2}}\right) ^{n-1}=\frac{(-1)^{n-1}}{\ell_{\text{eff}}^{2n-2}}
+(n-1)\,\frac{(-1)^{n-2}}{\ell_{\text{eff}}^{2n-4}} \,\frac{%
r_{0}^{2n-3}}{r^{2n-1}}+\cdots \,.  \label{expansion}
\end{equation}%

The remaining terms in the sum can be neglected as they vanish in the limit $%
r\rightarrow \infty $, in which the surface integral is evaluated.
Keeping all the coupling constants arbitrary, we obtain the divergent expression%
\begin{eqnarray}
E &=&\text{Vol}(S^{2n-2})\lim_{r\rightarrow \infty }\left[ C\left( \frac{r^{2n-1}}{\ell _{\text{eff}%
}^{2}}+\frac{\left( 2n-3\right) r_{0}^{2n-3}}{2}\right) \right.   \notag \\
&&+\left. \left( 2(2n-3)\gamma +n\,\alpha _{2n}(2n-2)!\,\frac{(-1)^{n-2}}{%
2\ell _{\text{eff}}^{2n-4}}\right) \right] \frac{(2n-2)r_{0}^{2n-3}}{\ell _{%
\text{eff}}^{2}}\,,
\end{eqnarray}
where we introduced the constant
\begin{equation*}
C=\frac{1}{\kappa }-\frac{2}{\ell _{\text{eff}}^{2}}\left( \rule{0pt}{15pt}%
(2n-1)\left( \alpha +2n\beta \right) +(2n-2)(2n-3)\gamma \right) +n\,\alpha
_{2n}(2n-2)!\frac{(-1)^{n-1}}{\ell _{\text{eff}}^{2n-2}}\,.
\end{equation*}
Now it is clear that fixing the value of $\alpha _{2n}$ as in Eq.(\ref%
{alppha}), which produces $C=0$ and regularizes the $r^{2n-1}$ term, makes
the gravitational energy finite.

Finally, writing $\gamma $ in terms of $\alpha $ and $\beta $ (as given in
Eq.(\ref{cosmconst})), the energy is
\begin{equation}
E=\left( -1+\frac{8\Lambda _{\text{eff}}\,\kappa }{(2n-2)^{2}}\left( 2n\beta +\alpha
\right) +\frac{2\Lambda _{0}}{\Lambda _{\text{eff}}}\right) \frac{(2n-2)r_{0}^{2n-3}%
}{4G_{2n}}\,,  \label{ADT}
\end{equation}
what is the correct result for black hole mass in QCG theory with $\Lambda _{0}=0$, cf. \cite{Deser:Tekin}, in our case generalized to include the bare cosmological constant $\Lambda _{0}$.

In the case when the metric component $g_{tt}$ in the ansatz is replaced by a more general function $-N^2(r)f^2(r)$ that respects the AdS asymptotics, the resulting formula for the energy still yields a finite result. Here, in order to compare with known results in the literature, we consider the case $N=1$.

The above discussion shows the effect of the Euler topological invariant added to the QCG action, which is similar to the one
in GR and Einstein-Gauss-Bonnet gravity actions. Namely, without such term, the gravitational
energy is both divergent and yields an incorrect value for the finite part in the mass computation. The addition of properly
fixed Euler coupling $\alpha_{2n}$ resolves both issues at the same time.

\section{Conclusions}

In this paper, we derived a general formula for the conserved charges for asymptotically AdS
black holes in Quadratic Curvature Gravity in $2n$ dimensions. The addition of topological terms to
the gravity action acts as a regulator at the level of the surface term, and yields finite Wald charges
in AdS without need of extra boundary terms. It remains an open question to know whether the method of adding Euler characteristics can also be implemented in order to compute conserved charges in sectors with other asymptotic behaviors, such Lifshitz black holes, AdS-waves, or spacetimes with weakened asymptotics in AdS.

In the past, the use of topological terms to regularize the conserved quantities and the Euclidean action of a given gravity theory has not only provided a powerful, yet simple computational tool to understand physical properties of black holes, but it has also unveiled remarkable features of the gravitational action in critical points of the parameter space in four dimensions such as Critical Gravity \cite{Lu:Pope}. Indeed, the addition of the Gauss-Bonnet term to the Critical Gravity action makes manifest the fact that the energy, the entropy and the action are zero for Einstein spaces \cite{Miskovic:2014zja, Anastasiou:2017rjf}. Implications of this procedure in the computation of holographic correlation functions have been explored in Ref.\cite{Anastasiou:2017mag}. We expect that the addition of Euler terms will make more clear certain properties of critical theories in higher dimensions too.

\[\]
Note added: While finishing our paper, the work \cite{Meng} appeared, in which the higher-dimensional extension of \cite{GORD} is also worked out.

\acknowledgments

This work was funded in parts by the NSF through grant PHY-1214302, Chilean FONDECYT grants N$^{\circ}$1170765 and N$^{\circ}$1190533, and the grant VRIEA-PUCV N$^{\circ}$123.764/2019. D.R.B. is supported by Becas Chile (CONICYT) PhD scholarship No. 72200301. G.G. is partially supported by CONICET through the grant PIP 1109 (2017).

\end{document}